# Virtual Histology with Photon Absorption Remote Sensing using a Cycle-Consistent Generative Adversarial Network with Weakly Registered Pairs

James E.D. Tweel[1,2,†], Benjamin R. Ecclestone[1,2,†], Marian Boktor[3],
James Alexander Tummon Simmons[1,2], Paul Fieguth[3], Parsin Haji Reza[1,*]

[1]PhotoMedicine Labs, Systems Design Engineering, University of Waterloo, Waterloo, Ontario, N2L 3G1, Canada
[2]illumiSonics, Inc., 22 King Street South, Suite 300, Waterloo, Ontario, N2J 1N8, Canada
[3]Vision and Image Processing Lab, Systems Design Engineering, University of Waterloo, Waterloo, Ontario N2L 3G1, Canada
*Corresponding author: phajireza@uwaterloo.ca
†Authors contributed equally to this work

___

**Abstract** – Modern histopathology relies on the microscopic examination of thin tissue sections stained with histochemical techniques, typically using brightfield or fluorescence microscopy. However, the staining of samples can permanently alter their chemistry and structure, meaning an individual tissue section must be prepared for each desired staining contrast. This not only consumes valuable tissue samples but also introduces delays in essential diagnostic timelines. In this work, virtual histochemical staining is developed using label-free photon absorption remote sensing (PARS) microscopy. We present a method that generates virtually stained histology images that are indistinguishable from the gold standard hematoxylin and eosin (H&E) staining. First, PARS label-free ultraviolet absorption images are captured directly within unstained tissue specimens. The radiative and non-radiative absorption images are then preprocessed, and virtually stained through the presented pathway. The preprocessing pipeline features a self-supervised Noise2Void denoising convolutional neural network (CNN) as well as a novel algorithm for pixel-level mechanical scanning error correction. These developments significantly enhance the recovery of sub-micron tissue structures, such as nucleoli location and chromatin distribution. Finally, we used a cycle-consistent generative adversarial network CycleGAN architecture to virtually stain the preprocessed PARS data. Virtual staining is applied to thin unstained sections of malignant human skin and breast tissue samples. Clinically relevant details are revealed, with comparable contrast and quality to gold standard H&E-stained images. This work represents a crucial step to deploying label-free microscopy as an alternative to standard histopathology techniques.

## I.   INTRODUCTION

Modern pathologists study the microscopic anatomy of tissue specimens to understand the nature and progression of disease. To perform microscopic inspection using brightfield or fluorescence microscopes, tissue specimens are thinly sectioned and stained with histochemical dyes. These dyes chemically label the structures and biomolecules within the sample, facilitating the differentiation of key tissue elements such as lipids, proteins, and nucleic acids [1]. The most prevalent stain set used in histology and cancer diagnosis is hematoxylin and eosin (H&E). Hematoxylin stains the chromatin in the nuclei purple, while eosin stains the cytoplasm and extracellular structures pink [2]. These contrasts enable pathologists to identify both tissue and nuclear abnormalities which indicate the presence, nature, and extent of malignancy. Depending on the disease, other specialized stains may be used to assess targeted tissue features. For example, Grocott's methenamine silver stain (GMS) or periodic acid Schiff (PAS) stains may be used to highlight fungal cells if a fungal infection is suspected [3]. In certain cases, advanced labelling techniques may be employed to identify specific proteins, or RNA/DNA sequences [4], [5]. These methods facilitate highly specific diagnostics, such as the identification of genetic subtypes in cancers. For example, immunohistochemical (IHC) staining or fluorescence *in situ* hybridization (FISH) are used to identify HER2 positive breast tumors [6], where HER2 specific treatments have significantly improved patient outcomes [7].

In practice, simultaneous or sequential use of histochemical, IHC, and FISH agents is not possible on a single tissue section. The labelling process can introduce irreversible structural and chemical changes which render the specimen unacceptable for subsequent analysis [2], [8]. As such, each section must be independently sectioned, mounted, and stained; a technically challenging, expensive, and time-consuming workflow [9]. A trained histotechnologist may spend several hours to prepare a section for testing [10], with some labeling protocols requiring overnight incubation and steps spaced out across multiple days [11]. Hence, repeating staining or producing additional stains in a stepwise fashion can delay diagnostics and treatment timelines, degrading patient outcomes. Moreover, performing multiple stained sections can rapidly expend invaluable diagnostic samples, particularly when the diagnostic material is derived from needle core biopsies. This increases the probability of needing the patient to undergo further procedures to collect additional biopsy samples, incurring diagnostic delays, and significant patient stress.

Label-free microscopy modalities offer an opportunity to revolutionize modern digital pathology by enhancing the diagnostic utility of valuable tissue specimens. Label-free microscopes leverage biomolecules endogenous optical characteristics to capture chromophore specific visualizations without histochemical labeling [12]. This opens the possibility of directly imaging within unprocessed tissue specimens, potentially facilitating in-vivo histological imaging in the future. When combined with deep-learning image translation techniques, label-free microscopes enable virtual histochemical staining from unlabelled tissue specimens. Ideally, label-free microscopy could provide pathologists immediate access to numerous specialized stains, enhancing diagnostic confidence while reducing processing time and tissue requirements. Towards this end, several modalities have recently achieved some success in developing deep-learning based label-free virtual histochemical staining including, quantitative phase imaging (QPI) [13], optical coherence tomography (OCT) [14], photoacoustic microscopy [15], and autofluorescence microscopy [16]–[18]. Additionally, multimodal non-linear microscopy techniques, such as coherent anti-Stokes Raman scattering, two-photon excitation fluorescence and second-harmonic generation, have been used for virtual H&E staining of tissue specimen [19].

These techniques have all shown some success in emulating one or more histochemical stains, however, their effectiveness primarily relies on the raw initial label-free contrast they can capture. Ideally, a given modality is able to recover sufficient chromophore specific data to match the desired chemical contrast, however this is not always the case. For example, while a sample's autofluorescence spectrum reveals significant information on its composition [20]–[22], some critical biomolecules may not possess distinct or measurable autofluorescent characteristics. While elastin, collagen, and other extranuclear constituents exhibit strong emissions, DNA and RNA have relatively low fluorescence quantum yield [23], which limits the measurement of nuclear contrast. As such, the staining network must predict nuclear contrast from surrounding structure as opposed to direct measurement. This, in turn, may limit the accuracy of histochemical staining emulation.

For recovery of direct nuclear contrast, a modality that utilizes non-radiative (e.g., photothermal, and photoacoustic) relaxation of biomolecules can be used [24]–[26]. One such technique, known as Photon Absorption Remote Sensing (PARS) and previously called Total-Absorption Photoacoustic Remote Sensing, is able to concurrently measure both the non-radiative and radiative relaxation processes [27]. By capturing both absorption fractions simultaneously, PARS is able to recover rich biomolecule specific contrast, such as quantum efficiency ratio, not afforded by other independent modalities. In PARS, the optical relaxation processes (radiative and non-radiative) are observed following a targeted excitation pulse incident on the sample [27]. The radiative relaxation generates optical emissions from the sample which are then directly measured. The non-radiative relaxation causes localized thermal modulations and, if the excitation event is sufficiently rapid, pressure modulations within the excited region. These transients induce nano-second scale variations in the sample's local optical properties, which are captured with a co-focused detection laser. Additionally, the co-focused detection is able to measure the local optical scattering prior to excitation. Overall, PARS is able to simultaneously capture radiative and non-radiative absorption as well as and optical scattering from a single excitation event.

Label-free virtual histology has previously been explored using an ultraviolet (UV, 266nm) excitation PARS platform [28]. UV excitation aligns with the absorption peak of several relevant biomolecules such as DNA, RNA, collagen, and elastin [29]. Important nuclear contrast comes primarily from the non-radiative relaxation of DNA, while surrounding connective tissue contrast comes from the radiative relaxation of extranuclear proteins. These combined PARS label-free contrasts are highly analogous to traditional chemical H&E staining. Recent work employed a pix2pix image translation network on PARS data for H&E emulation [28]. This supervised approach requires exact pixel-to-pixel matched ground truth data for emulation [30]. Perfect alignment of the datasets is not only challenging but is often not possible due to the deformations and potential degradations of the tissue specimen caused by the staining process. Misalignment in training pairs can significantly impact and compromise the quality of the pix2pix results.

We present an improved virtual staining and image processing workflow for emulating histology images which are effectively indistinguishable from gold standard H&E pathology. The presented developments include a new staining network, and an optimized image preprocessing pathway. Here, a cycle-consistent generative adversarial network (CycleGAN [31]) architecture is applied for virtual staining. CycleGAN virtual staining does not require pixel-to-pixel level registration for training data [31]. However, semi-registered data is used here to reduce hallucination artifacts [32], while improving virtual staining integrity. In addition, advances in image preprocessing reduce inter-measurement variability during signal acquisition. Improvements include pulse energy correction and image denoising using the self-supervised Noise2Void network [33]. Additionally, a novel algorithm is presented for removal of pixel level mechanical scanning position artifacts, which blurs subcellular level features. These enhancements afford marked improvements in the clarity of small tissue structures, such as nucleoli and chromatin distribution. Direct comparisons are made between the previous pix2pix, standard unpaired CycleGAN, and the proposed loosely registered CycleGAN virtual colourizations. The loosely registered CycleGAN facilitates precise virtual staining with the highest quality of any PARS virtual staining method explored to date. When applied to entire whole slide sections of resected human tissues, the proposed virtual staining provides detailed emulation of subcellular and subnuclear diagnostic features comparable to the gold standard H&E. This work represents a significant step towards the development of a label-free virtual staining microscope. The successful label-free virtual staining opens a pathway to the development of in-vivo virtual histology, which could allow pathologists to immediately access multiple specialized stains from a single slide, enhancing diagnostic confidence, improving timelines and patient outcomes.

## II. MATERIALS AND METHODS

### A. Sample Preparation

Tissue samples were first fixed in formalin solution for a period of 24 to 48 hours, within 20 minutes of excision. Samples were then dehydrated with ethanol and treated with xylene to eliminate residual ethanol and fats. The samples were subsequently embedded in paraffin wax, creating formalin-fixed paraffin-embedded (FFPE) blocks. A microtome was then used to cut thin tissue sections (~4-5μm) from the FFPE blocks. Tissue sections were placed on glass microscope slides and baked at 60°C for approximately 60 minutes to evaporate excess paraffin. The unstained samples were first imaged at 40x with the PARS microscope and then directly stained with H&E. H&E-stained slides were then imaged at 40x (Morpholens 1, Morphle Digital Pathology). This process was performed on a variety of malignant human skin and breast tissue samples and direct one-to-one whole slide images were acquired for model training.

Tissues were provided by clinical collaborators at the Cross-Cancer Institute (Edmonton, Alberta, Canada) from anonymous patient donors with all patient identification removed from the samples. Patient consent was waived by the ethics committee because these archival tissues were no longer required for patient diagnostics. No information regarding patient identity was provided to the researchers. Samples were collected under protocols approved by the Research Ethics Board of Alberta (Protocol ID: HREBA.CC-18-0277) and the University of Waterloo Health Research Ethics Committee (Photoacoustic Remote Sensing (PARS) Microscopy of Surgical Resection, Needle Biopsy, and Pathology Specimens; Protocol ID: 40275). All human tissue experiments were conducted in accordance with the government of Canada guidelines and regulations, including "Ethical Conduct for Research Involving Humans (TCPS 2)".

### B. Description of PARS Tissue Imaging Process

The label-free images acquired for this study were captured using the whole slide scanning PARS system previously reported by Tweel *et al.* [34]. In short, a 400ps pulsed 50KHz 266nm UV laser (Wedge XF 266, RPMC) is used to excite the sample, simultaneously inducing non-radiative and radiative relaxation processes. The non-radiative relaxation processes are sampled as time-resolved photothermal, and photoacoustic signals probed with a continuous wave 405nm detection beam (OBIS-LS405, Coherent). This detection beam is co-aligned and focused onto the sample with the excitation light using a 0.42 numerical aperture (NA) UV objective lens (NPAL-50-UV-YSTF, OptoSigma). The radiative emissions (>266nm) from the radiative relaxation process, as well as the transmitted detection light, are collected using a 0.7 NA objective lens (278-806-3, Mitutoyo). The 405nm detection wavelength and the radiative emissions are spectrally separated, and each directed toward an avalanche photodiode (APD130A2, Thorlabs).

To form an image, mechanical stages move the sample in an "s"-like scanning pattern to laterally separate the excitation events on the sample (~250nm/pixel). At each excitation pulse, several hundred nanoseconds of time-resolved signal from each system photodiode is digitized at a 200MHz rate (CSE1442, RZE-004-200, Gage Applied). A portion of the collected signal is pre-excitation and is used to form the scattering image of the sample in its unperturbed state. The non-radiative image pixels are then derived as a percentage modulation in the detection scattering (post-excitation). Next, the radiative image pixels are obtained from the peak emission amplitude recorded after each excitation event. Pixels are then arranged in a cartesian grid based on the stage position feedback, forming a stack of three co-registered label-free image contrasts: non-radiative, radiative, and scattering. Finally, the excitation pulse energy and detection power, recorded throughout imaging, are used to correct image noise caused by laser power and pulse energy variability.

Whole slide samples can be scanned using the automated workflow previously described by Tweel *et al.* [34]. In brief, the entire tissue area is divided into subsections (500x500µm), each individually scanned at their optimal focus position. Using their relative stage positions and small amount of overlap (~5%), these sections are stitched and blended into a single whole slide image.

### C.  *PARS Data Preprocessing*

In addition to the correction of noise due to laser power and pulse variability, the Noise2Void (N2V) framework developed by Krull et al. [33] is used to further denoise the raw PARS images. Unlike many other traditional CNN-based denoising methods, N2V does not require paired training data with both a noisy and clean image target. It assumes that image noise is pixel-wise independent, while the underlying image signal contains statistical dependencies. As such, it facilitates a simple approach for denoising PARS images, and was used to train a denoising CNN for the radiative and non-radiative contrast channels, separately. Models were trained on a body of raw data taken from both human skin and breast whole slide images. A series of 125 PARS tiles was used to generate a model for each of the radiative and non-radiative images. Each model was trained over a series of 300 epochs, with 500 steps per epoch, using 96 pixel neighbourhoods. The final processing step before training the virtual staining model is to correct a scanning-related image artifact, which is uncovered after denoising the raw data. These artifacts are line-by-line distortions caused by slight inconsistencies in the mechanical scanning fast axis (x-axis) velocity, which results in uneven spatial sampling. As such, before colourization a custom jitter correction algorithm is used to fix these distortions (see more information in Supplemental Information Section A).

### D.  *Dataset Preparation for Model Training*

In this work, a CycleGAN image translation model was used for virtual staining. While CycleGAN is able to learn an image domain mapping with unpaired data, it can be advantageous to provide the model with semi or loosely registered images, as a form of high-level labeling to better guide the training process and strengthen the model. As one-to-one H&E and PARS whole slide image pairs are obtainable, it seems most appropriate to prepare the dataset accordingly. However, the two datasets are not intrinsically registered, so a simple affine transform is used. Affine transforms allow for shearing and scaling, as well as rotation and translation [35]. In general, it is sufficient for the alterations of tissue layout on the slide which occur during the staining process. The affine transform is determined using the geometric relationship between three registration points. This found relation, or transformation matrix, is then applied to the entire whole slide image for both the non-radiative and radiative channels [35].

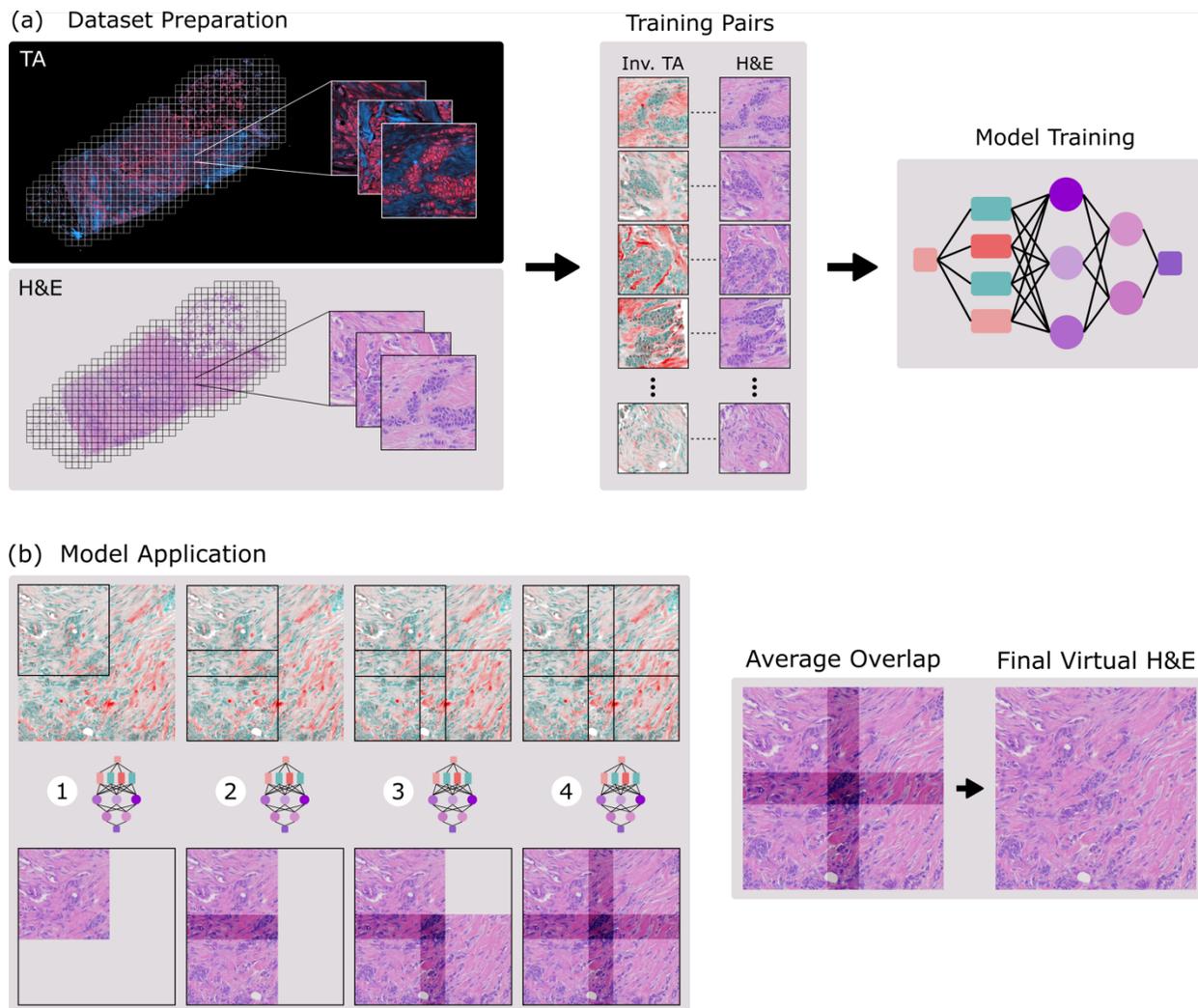

Figure 1: Visualization of data preparation process and inversion. (a) The registered total-absorption and H&E images are cut into matching tiles, to generate a loosely registered dataset. The pixel intensities of the total-absorption images are then inverted, to provide a better initialization for training. Finally, the datasets are used to train the virtual colorization model. (b) To form images, the model is repeatedly applied to overlapping tiles of the total absorption images. The overlapping tiles are subsequently averaged to form the final virtual colorization.

After the whole slide PARS Total Absorption (TA) image and H&E image are registered, the entire image is sliced into small tiles (512x512) which are paired together as shown in Figure 1(a). The total absorption (TA) image shows the radiative (blue) and non-radiative (red) raw images in a combine single colored image. However, during training, the network uses inverted TA patches, in which the radiative and non-radiative image pixel intensities are inverted before they are stacked into a colored image. Inverting these channels provides a colored image where the white background in the PARS data maps to the white background in the H&E data. After training is complete, the model can be applied to larger images, such as entire whole slide images, by virtually staining 512x512 tiles in parts. This process is shown in Figure 1(b) for a smaller inverted TA image. When applying the model, the virtually stained tiles overlap, and these overlap regions are averaged together in the final virtually stained image. Here an overlap of 50% was used.

In this study two CycleGAN models were trained on loosely paired data using the registration and dataset preparation methods described earlier. One model was trained on human skin tissue and another on human breast tissue. For each model, the training sets were composed of 5000 training pairs of size 512x512px (128x128 μm) sourced from standard 40x magnification (250nm/pixel) whole slide images of each tissue type. The model generators were trained for 500 epochs with an early stopping criteria to terminate training when losses stopped improving. The model was trained with a learning rate of 0.0002, batch size of 1 and an 80/20% split of training and validation pairs. For comparison purposes, a pix2pix model and standard unpaired CycleGAN model were also trained for each tissue type. The pix2pix models were trained on the same dataset as the paired CycleGAN model, however with the more rigorous registration process and the same model parameters previously described by Boktor *et al.* [28]. For the unpaired training of CycleGAN models, the same number of training pairs were used, however the TA and H&E domains were sourced from different whole slide images of the same tissue type.

## III. RESULTS AND DISCUSSION

A current shortcoming of the PARS raw images is the presence of measurement noise. In a recent work by Tweel *et al.* [34] significant improvements in PARS image quality were achieved by measuring detection power and excitation pulse energy. Image noise was then correction based on the laser energy variability. Even with the energy reference correction, measurement noise is still present in the non-radiative signals. This additive noise disproportionately impacts signals which exhibit low non-radiative relaxation since they generate smaller non-radiative perturbations in the detection beam. Figure 2 shows an example of the raw non-radiative and radiative image channels after reconstruction and laser power reference correction. At high magnification, significant noise can be seen in the raw data channels. This motivates denoising as a preprocessing step. However, noiseless PARS image targets were not available for training a traditional denoising CNN. Hence, the N2V framework, described in section II.C is an ideal method as it allows effective denoising without a clean image target. Figure 2 shows results after denoising with clear improvements in image quality for both the non-radiative and radiative channels. After removing noise from the raw data, the jitter artifacts mentioned in Section II.C are uncovered and become the main source of noise in the images. While these sub-resolution shifts and distortions between the rows of the image can be seen embedded within the noise, they are difficult to resolve and correct. Denoising not only helps improve raw data quality but helps make the jitter correction possible. As shown in Figure 2, most of the artifacts are removed after applying the jitter correction algorithm (more information in Supplemental Information Section A).

After denoising and jitter correcting the raw data, the whole slide radiative and non-radiative images are registered to the ground truth H&E image. As mentioned in Section II.D a simple affine transform is used here to account for the tissue layout alterations accrued during the staining process. The three-point affine registration is less rigorous compared to the methods employed by Boktor *et al.* [28] for pix2pix virtual staining. However, it is significantly faster and may generate upwards of 6000 closely registered 512x512 training pairs for a single 40x, 1cm$^2$, whole slide image. An example of a whole slide image before and after registration can be seen in Supplemental Information Section B (Figure S2) for a breast tissue sample.

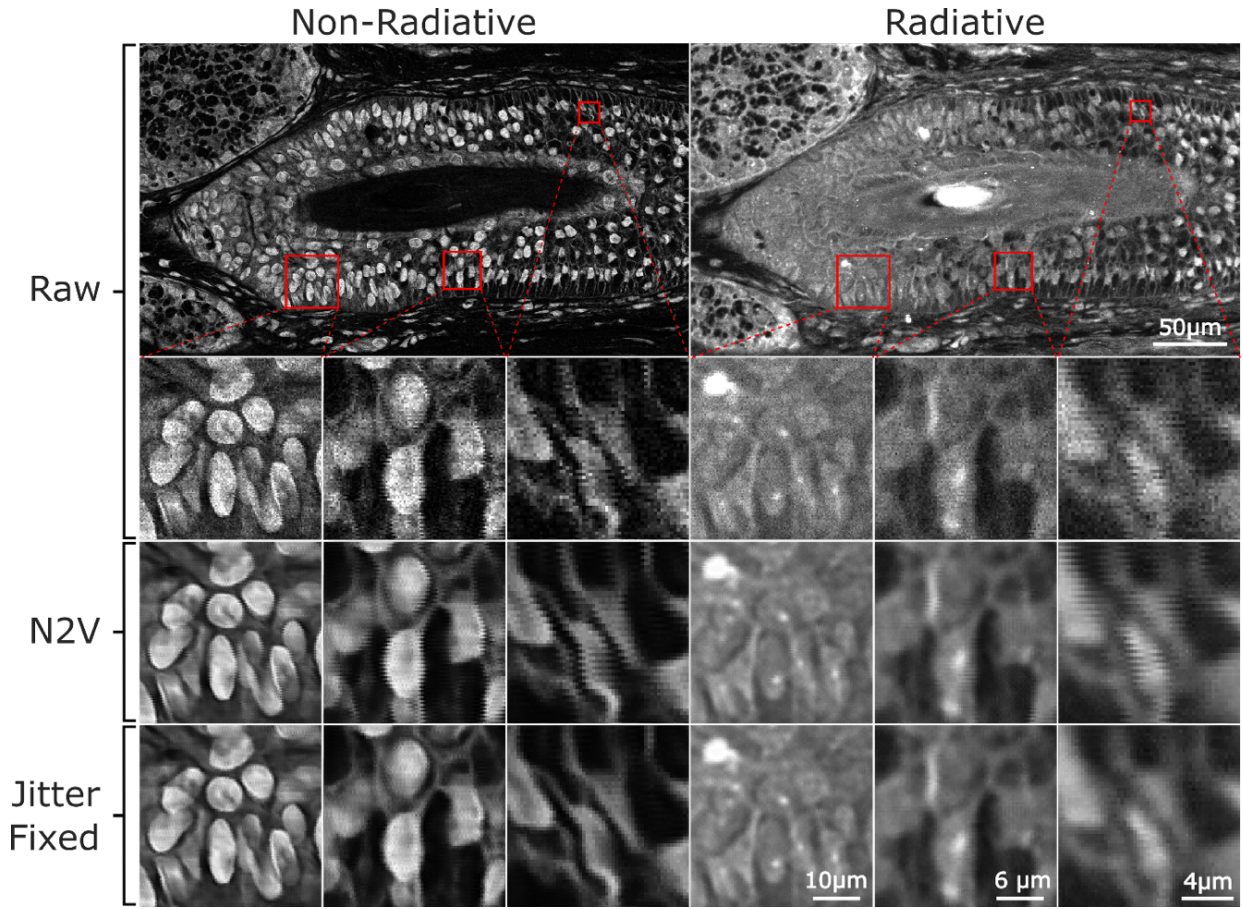

Figure 2: Denoising results with the N2V-based denoising CNN and subsequent jitter correction algorithm applied to both the raw non-radiative and radiative image channels. Three example regions are shown at higher magnification to see the effect of the denoising and jitter correction algorithms. The structure imaged here shows a hair follicle capture from human skin tissue.

In some cases, alignment between these training pairs may be sufficient for error metrics specific to supervised learning, such as means squared error (MSE), or structural similarity index measure (SSIM). In the future this may enable hybrid training schemes which combine paired and unpaired data. Previous works in hybrid image-to-image translation have shown improve performance over purely unsupervised methods, even with a small amount of additional paired data [36], [37]. Although supervised error metrics were not explicitly employed on the loosely paired CycleGAN presented here, the results demonstrate a potential advantage over strictly supervised pix2pix, and over CycleGAN trained entirely on unpaired data.

Next, a comparison between previous pix2pix based colorizations [28], unpaired CycleGAN, and the proposed paired CycleGAN was conducted. Figure 3 shows this comparison for unseen data from a variety of skin tissue structures. The pix2pix model was trained on the same PARS images used for the paired CycleGAN, however, registration was performed according to the previously reported process [28]. While the pix2pix model performs quite well, the transform tends to blur very fine structures. This artifact is observed in all four skin tissue examples. This blurring is likely caused by slight imperfections in alignment and registration between the training H&E and TA images, which severely weaken model performance.

Achieving perfect alignment between the label-free images and H&E data is challenging as certain structures are very susceptible to staining artifacts. For example, lipid areas or regions with loose connective tissue may be severely altered, or even washed away during the deparaffinization and staining processes. This leads to inconsistencies between datasets which cannot be accounted for through registration. Subsequently, even if the label-free and H&E-stained images are aligned, there may still be variation or uncertainty in the data which affect translation quality. This highlights the importance of a more flexible model such as CycleGAN which can handle variation and uncertainty in the input data.

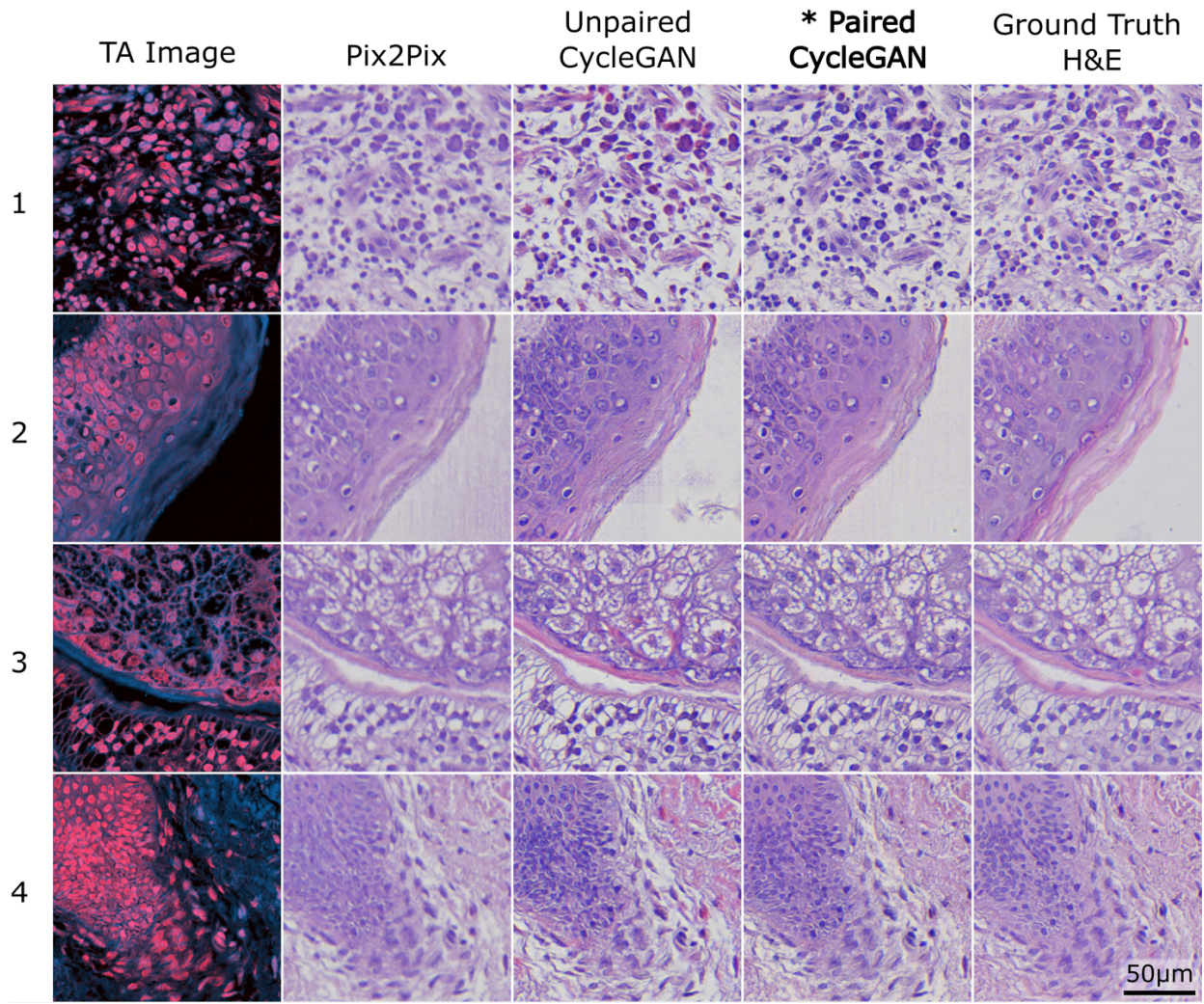

\* Proposed Method

Figure 3: Comparison of pix2pix as well as paired and unpaired CycleGAN implementation with the gold standard H&E and PARS total absorption (TA) acquisitions of skin tissue.

As expected, both unpaired and paired CycleGAN implementations provide sharper virtual H&E (Figure 3). However, the model trained on an unpaired dataset shows some examples of mis-colorization. For example, Figure 3 Section 1 shows a cluster of cell nuclei which, in the unpaired model, have been tainted red and look to be colorized as red blood cells instead of cell nuclei. Conversely, Section 2 shows a crop of the skin's epidermis layer with some background whitespace. The unpaired CycleGAN adds texture to the outermost stratum coronium layer of the tissues. In addition, a hallucination structure is seen in the bottom right of the background, which is not seen in the TA image, or in the H&E. In Section 3, which shows part of a sebaceous gland, the unpaired CycleGAN overemphasizes the red colouration in certain regions. In Section 4, a similar overemphasis of red colors is observed. The red incorrectly implies the presence of red blood cells, in the connective tissue. In unpaired scenarios, CycleGAN is a highly under-constrained model, which has been known to create hallucinations. Hallucinations occur when spurious structures are added or features are removed during the image domain transfer [32]. This problem usually arises when the data provided in the target domain has under or over representation, or bias, towards certain image features. Providing the CycleGAN model with loosely paired training data can strengthen the model during training by ensuring an equal representation of features in both image domains (TA and H&E). Hence, in all four diverse tissue structure examples, the paired CycleGAN does not exhibit the same artifacts as the unpaired version. It is clear the CycleGAN implementation trained with paired data has superior performance, avoiding the hallucination of structures seen with the unpaired CycleGAN model.

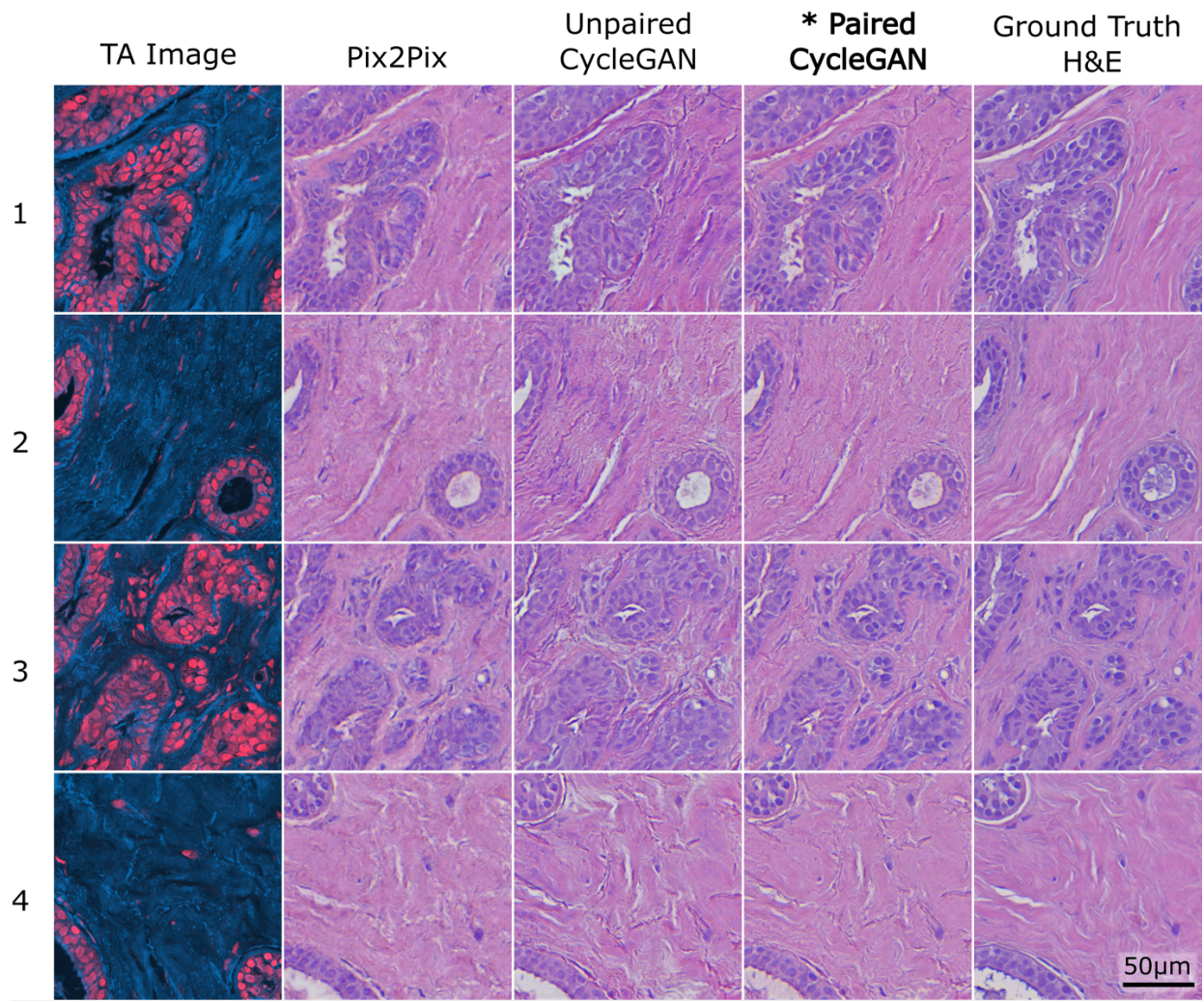

\* Proposed Method

Figure 4: Comparison of pix2pix as well as paired and unpaired CycleGAN implementation with the gold standard H&E and PARS total absorption (TA) acquisitions of malignant breast tissue.

The same comparison between colorization models was also performed on unseen malignant breast tissue areas, showing mainly glandular structures and connective tissues. These comparisons can be seen in Figure 4. As with the skin tissue examples, the data used for training was taken from separate whole slide images of the same tissue type. Overall, similar trends in the model performances are seen in the breast tissue structures. The pix2pix model applies a slight blur during the transformation and important nuclear details appear smudged. The unpaired CycleGAN implementation again produces hallucinations in the translation process. In all four examples, these hallucinations cause mis-colourization of the hematoxylin stain. Emulated hematoxylin stain is incorrectly spattered across areas of the connective tissue, falsely indicating the potential presence of nuclei. Abnormal nuclear morphology, such as increased nuclear size and irregular shape, organization, and patterning are all valuable details in cancer diagnosis and prognosis. Hence, these nuclear hallucinations are problematic. In contrast, the paired CycleGAN implementation avoids these hallucinations, and the virtual H&E closely resembles the ground truth.

Whole slide images provide pathologists with critical access to both low and high magnifications of tissue structures. This enables them informed diagnostic decisions. At low magnification, pathologists can overview the tissue structure, identify areas of concern, and contextualize high magnification analysis. High magnifications allow for in-depth examination of tissue structure and cellular morphology which is vital for precise diagnoses. As such, demonstrating virtual staining on an entire whole slide image is an essential goal. In Figure 5 the semi-registered CycleGAN virtual staining model is applied to an entire malignant skin tissue sample. This tissue contains a diverse set of structures to assess the model's performance. Both low and high magnification images are shown and compared against the corresponding one-to-one H&E whole slide image. Figure 5(a) shows a low magnification depiction of the entire slide where a variety of structures can be identified including an artery, smooth muscle structures, and a basaloid tumor extending from the epidermis. This indicates a primary diagnosis of basal cell carcinoma for the skin sample.

Figure 5(b) shows virtually stained sebaceous gland and hair follicle structures, stemming from the epidermis layer of the tissue. The virtual stain closely resembles the gold standard H&E. However, in the ground truth H&E, a blue color can be seen at the edge of the epidermis which is not shown in the virtually stained image. This blue inking is an artifact of the markings drawn on the tissue specimen by a surgeon during the resection process. Furthermore, part of the connective structures inside the void in the hair follicle bulb have been washed away during the deparaffinization and staining process. A similar removal of connective tissue can be in the upper right corner of the H&E image. As well, a few of the secretory cells in the sebaceous gland have been removed and appear fragmented. Notably, in Figure 5(d) the H&E image exhibits slash marks which are not present in the virtual H&E. In contrast, all these damaged structures are intact in the PARS virtual H&E. These artifacts exhibit why perfect registration is not always possible in certain parts of the tissue specimen. Additionally, they are prime examples of artifacts which may occur during histochemical processing. A potential advantage of implementing virtual staining is reducing such sample processing artifacts.

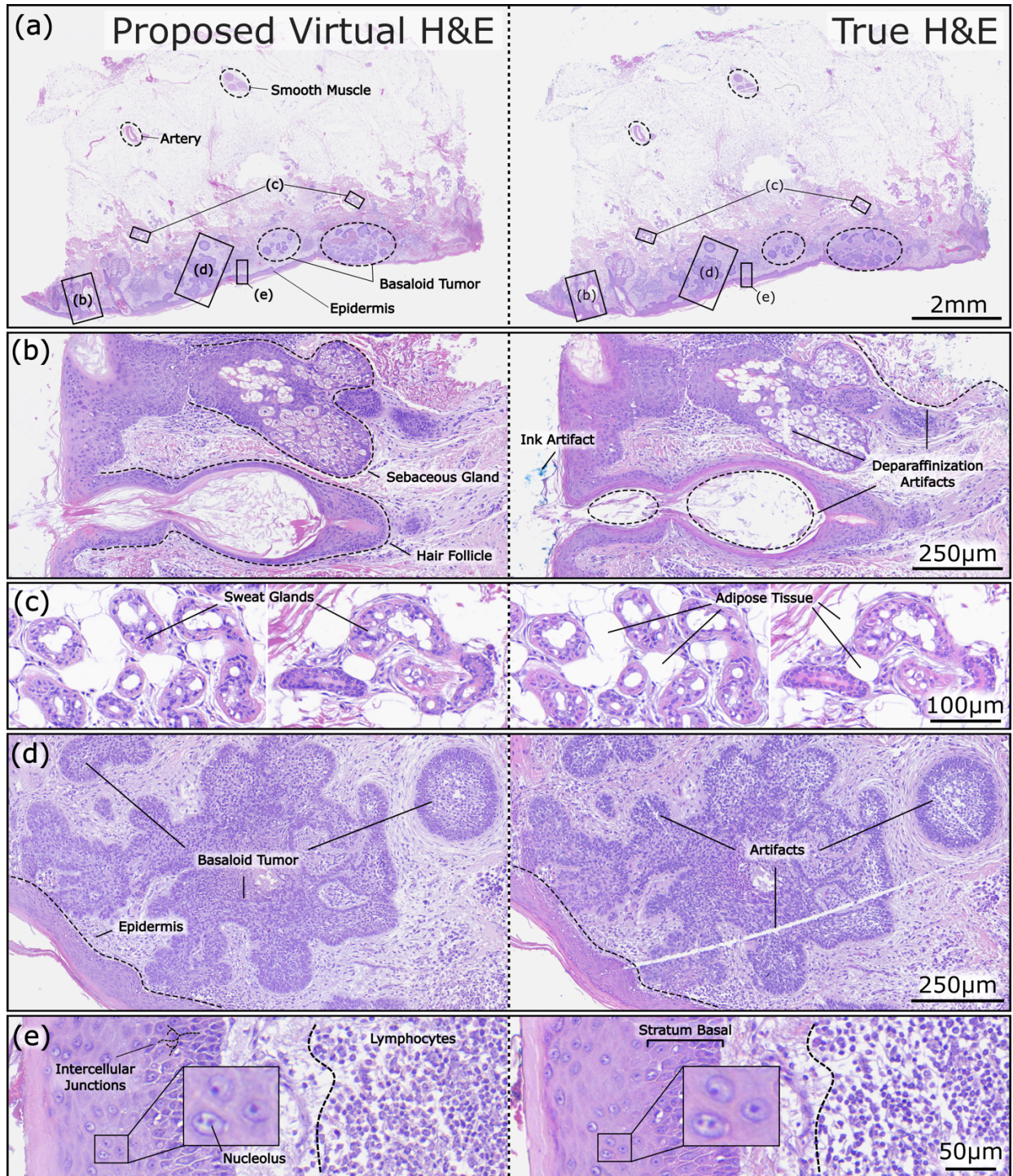

Figure 5: Comparison of the proposed paired CycleGAN virtual staining performance with the gold standard H&E on a sample of malignant skin tissue. (a) Whole slide visualization showing the epidermis, basaloid tumour, artery, and smooth muscles (b) High magnification view of a sebaceous gland and hair follicle. (c) High magnification visualization of sweat glands and adipose tissue. (d) Higher magnification view of the basaloid cells and basaloid tumor. (e) High magnification view of the epidermis, highlighting lymphocytes as well as detailed view of subcellar structures including nucleoli and intercellular junctions.

Figure 5(c) also shows the excellent performance of the semi-registered CycleGAN virtual staining model on a group of sweat glands. Here, the cytoplasmic membrane of the surrounding adipose tissue remains intact, and the lipid cell nuclei are recovered with clarity and color resembling the gold standard ground truth. Figure 5(d) shows a higher magnification view of the basaloid tumor which stems from the epidermis layer of the skin. The virtual H&E accurately mimics the staining colour of the basaloid cell nests of the tumour, which are important clinical features for diagnosis of basal cell carcinoma. Lastly, Figure 4(e) shows a close-up of the epidermal layers. Important diagnostic details can be seen in both the virtual and real H&E image, including subcellar structures such as nucleoli. Furthermore, within the stratum basale of the epidermis, a distinct network of thin lines and gaps is observed between the densely packed cells. These spaces are known as intercellular junctions which form intricate and essential networks of connections within the epidermis. These junctions help promote cell adhesion, facilitate intercellular communication, and maintain structural integrity in the epidermis. Proper observation of these structures is crucial during diagnostic evaluations. Disruptions or disorganization in these junctions can be indicative of cancer and potential invasiveness. Therefore, preserving the sharpness and quality of the input data, along with accurate staining colour, is of utmost importance in the virtual staining model. The paired CycleGAN model properly colourizes these smaller features and retains the input resolution of the raw data. This ensures that pathologists can effectively examine and evaluate these crucial aspects, leading to more accurate and reliable diagnoses.

As initially reported by Ecclestone *et al.* [27], the non-radiative and radiative label-free PARS contrasts tend to match the chemical staining contrast of hematoxylin and eosin respectively. However, there can still be notable distinctions in the chemical H&E staining and PARS visualizations. PARS may recover additional details and emphasize structures which are not highlighted by traditional H&E stains. In an example shown in Figure 6, the raw PARS image highlights the inner wall of an artery. Specifically, the internal elastic membrane, which provides elasticity and support to the artery, is emphasized in blue (radiative channel). This structure can be visualized with UV excitation due to the presence of certain fluorophores contained in the elastin fibres, one of which is a cross-linking tricarboxylic amino acid with a pyridinium ring [38]. In contrast, H&E is unable to show as clear a distinction between the layers of the artery. As such, PARS provides greater specificity to this structure compared to H&E. However, to accurately reflect the staining patterns of H&E, the virtual staining model deliberately supresses the intrinsic contrast to minimize the prominence of the internal elastic membrane.

Traditionally a stain such as Verhoeff-Van Gieson (VVG), which highlights normal or pathologic elastic fibers, would be required to visualize the internal elastic membrane of arteries [39]. In clinical applications, VVG stain is sometimes combined with Masson's trichrome stain [40], to differentiate collagen and muscle fibers within tissue samples. This is performed to visualize potential increases in collagen associated with diseases like cirrhosis and assess muscle tissue morphology for pathological conditions affecting muscle fibers. In contrast, all these structures are well highlighted in the PARS raw data. Currently, the H&E virtual staining model flattens these structures during the image translation process. However, this highlights the potential use of the rich PARS raw data to replicate various clinically relevant contrasts beyond H&E staining. Currently, clinical studies are underway to explicitly validate PARS virtual H&E as diagnostically comparable to chemical H&E visualizations. To this end, additional staining contrast will be explored in the near future. A primary goal for PARS virtual staining is to provide several emulated histochemical stains from a single acquisition. Moreover, there is a potential to develop completely new histochemical like contrasts based on the endogenous PARS contrast. PARS may be able to provide contrast to biomolecules which are inaccessible with current chemical staining methods.

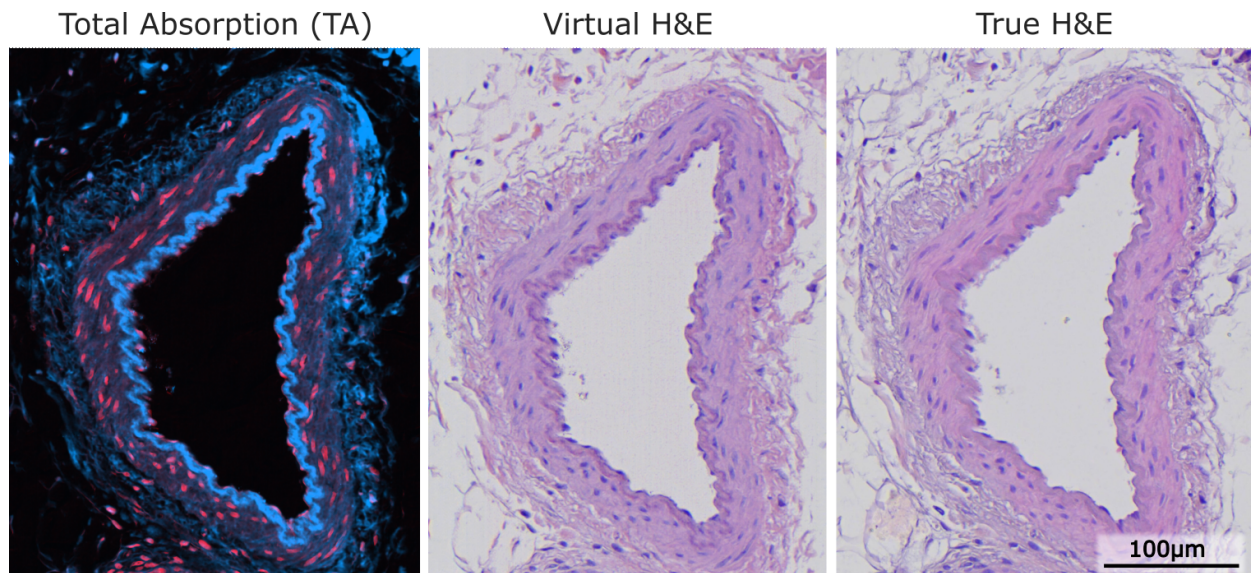

Figure 6: Example of the differences in the intrinsic PARS contrast, and chemical H&E staining. The PARS total absorption image highlights the inner wall of the artery or the internal elastic membrane. This feature is not highlighted in the H&E image.

## IV. CONCLUSION

We present an optimized PARS data processing and virtual staining method. Specific signal processing advances are exhibited which help to reduce measurement variability. Here, measurement reference correction, and Noise2Void based image denoising, are successfully applied to improve image quality. Finally, a new algorithm is presented to reduce pixel level mechanical scanning position artifacts, which blur submicron scale features. These enhancements afford marked improvements in the clarity of small tissue structures, such as nucleoli and chromatin distribution. In conjunction, a new virtual staining processes is presented which uses a semi-registered CycleGAN. While the semi-registered CycleGAN does not require registration like pix2pix, providing the semi-registered data may enhances the colorization quality by reducing the presence of hallucination artifacts. Presented here, emulated H&E images are produced from label-free PARS images with quality and contrast that compare favorably to traditional H&E staining. The colorization performance represents the current best PARS virtual staining implementation. Applied to entire sections of unstained human tissues, the presented method enables accurate recovery of subtle structural and subnuclear details. With these improvements, the PARS virtual H&E images, may be effectively indistinguishable from gold standard chemically stained H&E scans. This represents an essential milestone in developing a new clinically ready label-free virtual staining microscope. In the near future, PARS label-free virtual staining the has potential to provide multiple histochemical stains from a single unlabelled sample enhancing diagnostic confidence, and greatly improving patient outcomes.

**Funding**
The authors thank the following sources for funding used during this project. Natural Sciences and Engineering Research Council of Canada (DGECR-2019-00143, RGPIN2019- 06134); Canada Foundation for Innovation (JELF #38000); Mitacs Accelerate (IT13594); University of Waterloo Startup funds; Centre for Bioengineering and Biotechnology (CBB Seed fund); illumiSonics Inc (SRA #083181); New frontiers in research fund – exploration (NFRFE-2019-01012); The Canadian Institutes of Health Research (CIHR PJT 185984); NSERC Discovery Horizons DH-2023-00371.

**Acknowledgements**
The authors would like to thank Dr. Ally-Khan Somani, Dr. Gilbert Bigras and the Cross-Cancer Institute in Edmonton, Alberta for providing human breast and skin tissue samples. The authors would like to thank Hager Gaouda for helping prepare and stain the tissue samples used in this study. The authors would also like to thank Dr. John Mackey and Dr. Deepak Dinakaran for their help in clinical consultation in the assessment of the results.

**Author Contribution Statement**
Authors J.E.D.T and B.R.E contributed equally to this work.

**Competing Interests**
Authors James Tweel, Benjamin Ecclestone, James Alexander Tummon Simmons and Parsin Haji Reza all have financial interests in IllumiSonics which has provided funding to the PhotoMedicine Labs. Authors Marian Boktor and Paul Fieguth do not have any competing interests.

**SUPLEMENTAL INFORMATION**

***Section A: Custom algorithm for correction of scanning-related image jitter artifacts***

When the data stream acquired during scanning is shaped into an image section, pixels are evenly spaced in the cartesian grid. In reality, the spatial sampling is not perfectly even, owing to slight inconsistencies in stage velocity. This results in "jitter" artifacts which are then corrected using the custom algorithm shown in Figure S1. The artifacts and images shown in Figure S1 are exaggerated for the purpose of demonstrating the algorithm's working principles. The algorithm determines the optimal pixel shifts for a series of chunks spaced across a given row, with overlap. Chunks are then moved to their appropriate locations and summed together into a corrected row, with areas of overlapping chunks averaged. Figure S1 illustrates three example chunks and their optimal pixel shifts. These shifts are determined by moving a chunk left and right until a minimal mean square error is reached between the chunk and a reference row. This reference is calculated as the average between the top and bottom rows for the given row being corrected. The algorithm relies on the assumption that the fast axis speed profile differs mostly for velocity sweeps in opposing directions and minimally for velocity sweeps in matching directions. As such, the top and bottom rows were captured in the same direction and averaging them together provides a suitable in-between row to use as reference for correction.

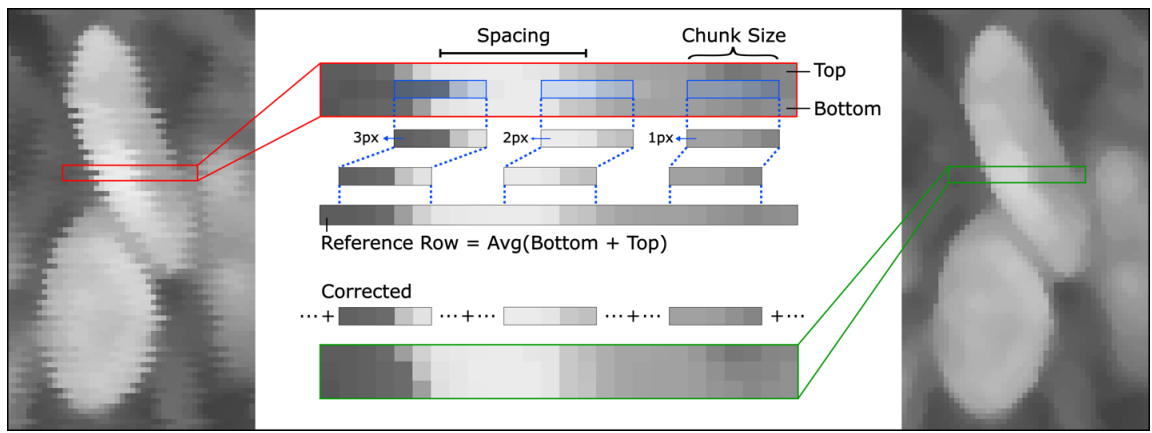

Figure S1: Algorithm used for jitter correction of denoised images. Each line of the image is broken into subsegments, which are correlated to the row above and below. Based on the correlation, the segment is shifted laterally to align the edges of image structures. Shifted segments are then averaged together to form a corrected line.

*Section B: Registration of a whole slide image using a three point affine transform*
Registration results are shown in Figure S2 for a small size breast tissue sample with an inverted TA image overlayed with ground truth H&E image. A close alignment can be seen between the PARS and H&E ground truth images at both low and high magnifications.

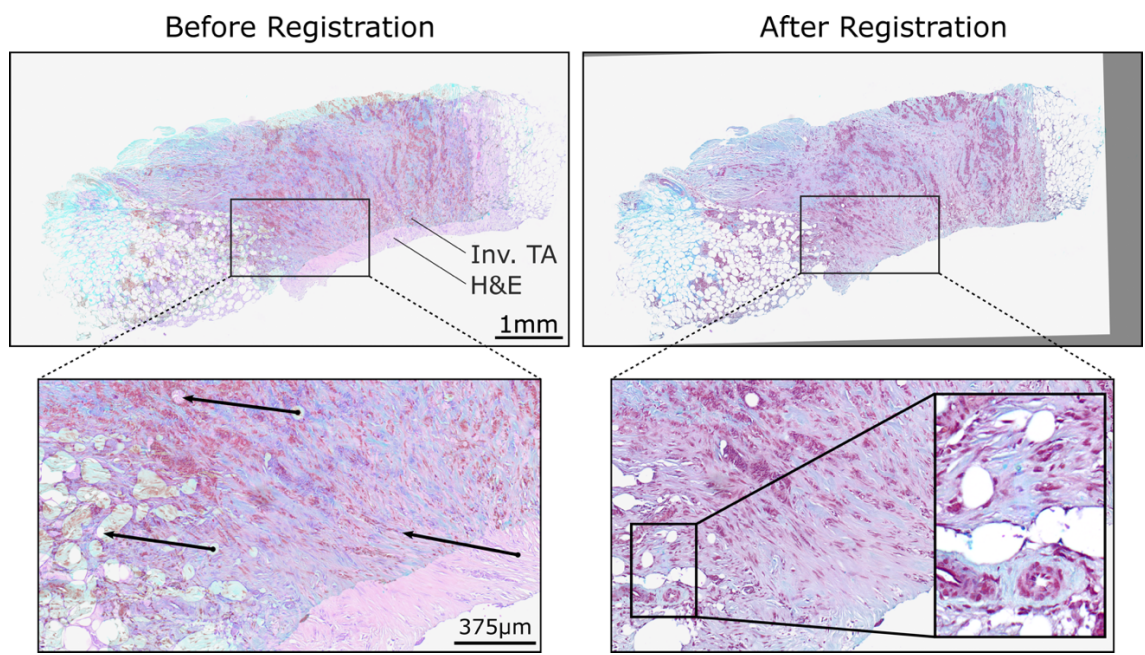

Figure S2: Example showing the before and after results of a three-point affine transform applied to an entire whole slide image. Three example points are shown as well as a high magnification close-up of the registration performance. Here the H&E and inverted TA images are overlayed.